\documentclass[aps,prl,amsmath,amssymb,floatfix,twocolumn,amsmath,superscriptaddress,twocolumn,nofootinbib,tighten,letterpaper]{revtex4-2}
\usepackage[colorlinks,linkcolor=blue,citecolor=blue,urlcolor=blue]{hyperref}
\usepackage{multirow}
\usepackage{subfigure}
\usepackage{color}
\usepackage{mathrsfs}
\usepackage{hyperref}
\usepackage[normalem]{ulem}
\usepackage{bm}

\usepackage{amssymb}
\usepackage{amsmath}
\renewcommand\vec[1]{\ensuremath\boldsymbol{#1}}

\usepackage{tabularray}

\usepackage{array} 
\newcolumntype{P}[1]{>{\centering\arraybackslash}p{#1}}

\definecolor{RowColor}{rgb}{0.88,1,0.9}

\usepackage{amsfonts, relsize, color, mathtools, physics}
\usepackage{graphics}
\usepackage{graphicx}
\usepackage{subfigure}
\usepackage{color}
\usepackage{comment}

\begin{document}

\title{Pair density wave in quarter metals from a repulsive fermionic interaction in graphene heterostructures: A renormalization group study}

\author{Sk Asrap Murshed}
\affiliation{Department of Physics, Lehigh University, Bethlehem, Pennsylvania, 18015, USA}

\author{Bitan Roy}
\affiliation{Department of Physics, Lehigh University, Bethlehem, Pennsylvania, 18015, USA}

\date{\today}

\begin{abstract}
Electronic bands in chirally stacked $n$ layer carbon-based honeycomb heterostructures, encompassing rhombohedral or ABC ($n \geq 3$), Bernal or AB bilayer ($n=2$), and monolayer ($n=1$) graphene, possess four-fold valley and spin degeneracy. Such systems with $n \geq 2$, when subject to external perpendicular electric displacement fields, feature a fully degenerate metal at high doping, a spin polarized but valley degenerate half-metal at moderate doping, and a non-degenerate quarter metal at low doping. Due to the fully polarized nature of the quasiparticles in the quarter metal, realized around one particular valley otherwise chosen spontaneously, it can sustain a single local superconducting ground state, representing a pair density wave that is chiral and odd parity in nature. From a leading order renormalization group analysis, here we show that repulsive density-density interaction among such polarized fermionic excitations can foster the pair density wave phase at low temperatures. Connections with experimentally observed superconducting states in the close vicinity of the quarter metal in some members of such graphene heterostructures family are discussed and possible routes to realize such a paired state in optical honeycomb lattices are highlighted. 
\end{abstract}

\maketitle

\emph{Motivation}.~Valley and spin polarized quasiparticles in the quarter metal of chirally stacked $n$ layer graphene heterostructure can foster a unique local or momentum-independent superconducting ground state (due to the Pauli exclusion principle) that is pair density wave (PDW) in nature with a characteristic $2{\bf K}$ periodicity where $\pm {\bf K}$ are the valley momenta~\cite{PDWgraphene:1, PDWgraphene:2}. Such a family of carbon-based honeycomb multilayers encompasses rhombohedral or ABC-stacked ($n \geq 3$), Bernal  or AB-stacked bilayer ($n=2$), and monolayer ($n=1$) graphene. Possibly, such a paired state has recently been observed in experiments in rhombohedral tetralayer ($n=4$) and hexalayer ($n=6$) graphene in the close proximity to the quarter metal in their global phase diagram in the $(D,n_e)$ plane, where $D$ is the perpendicular electric displacement field applied in the stacking direction and $n_e$ is the carrier density~\cite{PDWExp:1, PDWExp:2, PDWExp:3}. These experimental findings have triggered a new surge of theoretical works, exploring the unconventional properties of such observed superconductivity~\cite{ref:TTLGnew:1, ref:TTLGnew:2, ref:TTLGnew:3, ref:TTLGnew:4, ref:TTLGnew:5, ref:TTLGnew:6, ref:TTLGnew:7, ref:TTLGnew:8, ref:TTLGnew:9}. However, a clear microscopic origin of the PDW still remains unknown.

\begin{figure}[t!]
    \includegraphics[width=0.95\linewidth]{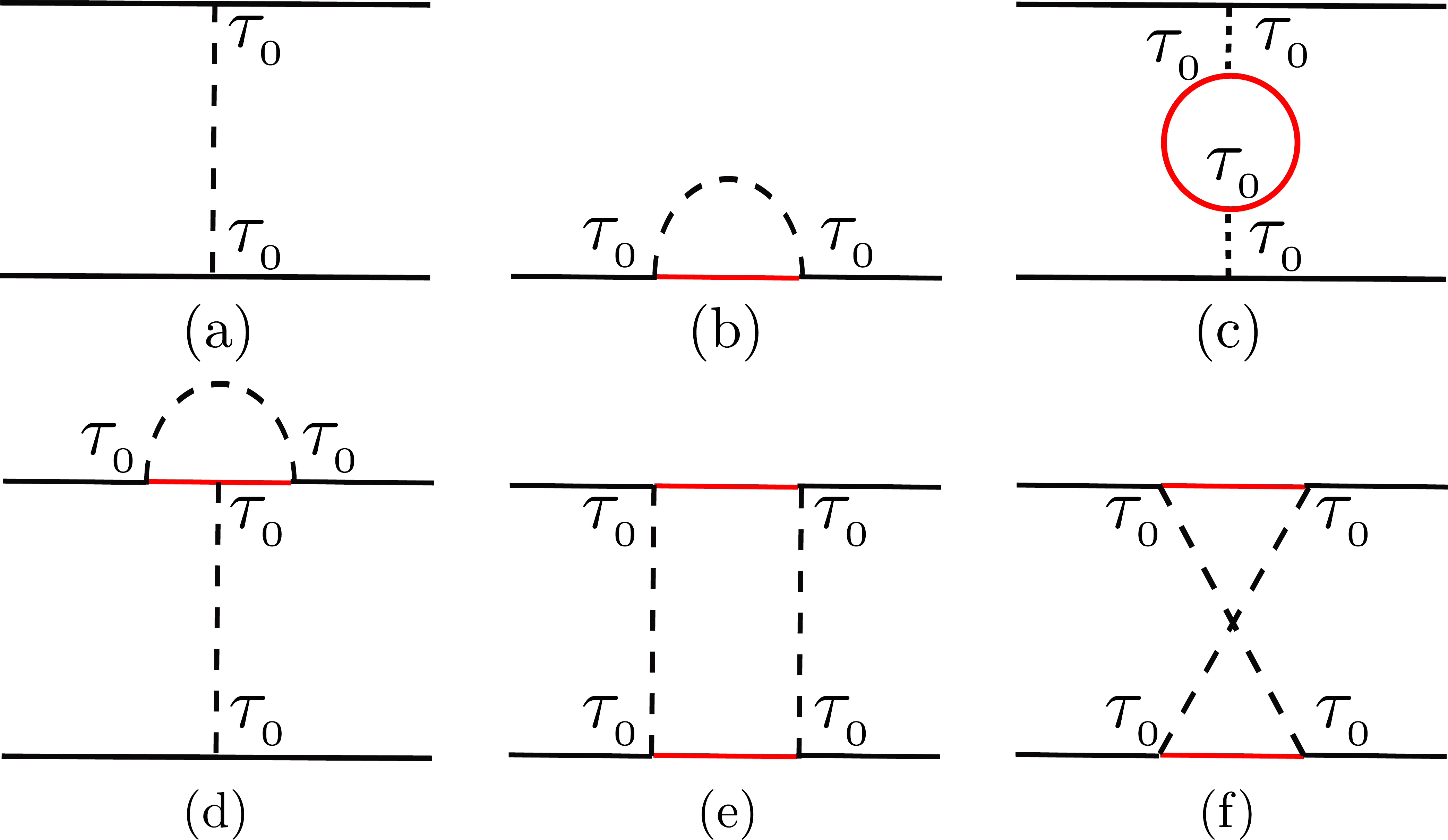}
    \caption{Feynman diagrams for (a) the bare four-fermion interaction $\left(\psi^{\dagger} \tau_0 \psi\right)^2$ and (b) the fermionic self energy correction, whose contribution yielding a renormalization of the chemical potential $\mu$ is, however, ignored within the framework of a leading-order renormalization group (RG) analysis. (c)-(f) Leading order corrections to the four-fermion interaction stem from the Feynman diagrams. Solid (dashed) lines represent fermions (interaction mediating bosons).  The red (black) solid lines represent fast (slow) modes with momentum $\Lambda e^{-\ell} < |\vec{k}| < \Lambda$ ($|{\vec k}| < \Lambda e^{-\ell}$), where $\Lambda$ is the ultraviolet momentum cut-off and $\ell$ is the logarithm of the RG scale.
    }~\label{fig:Feynman_Interaction}
\end{figure}

\begin{figure}[t!]
    \includegraphics[width=0.95\linewidth]{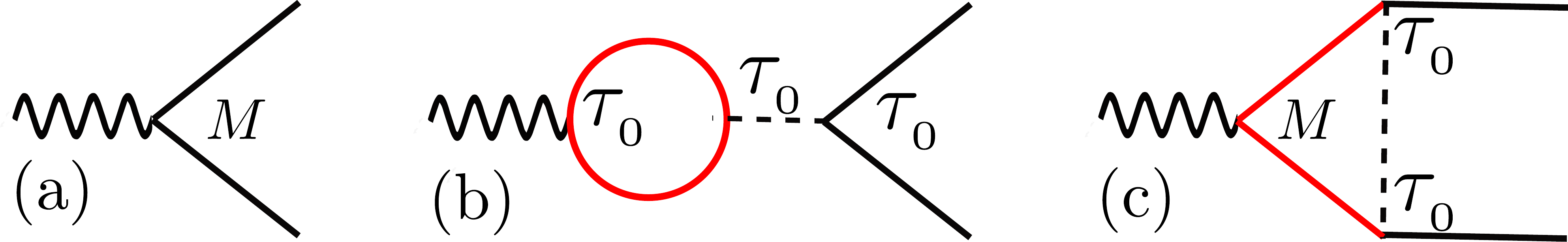}
    \caption{Feynman diagrams representing (a) the bare vertex for the source term $\psi^{\dagger} M \psi$ (excitonic) or $\psi^{\dagger} M \psi^\star$ (pairing), where $M$ is a Hermitian matrix, and (b) and (c) its leading order corrections. Rest of the details are the same as in Fig.~\ref{fig:Feynman_Interaction}.   
    }~\label{fig:Feynman_Source}
\end{figure}

\begin{figure*}[t!]
    \includegraphics[width=1.00\linewidth]{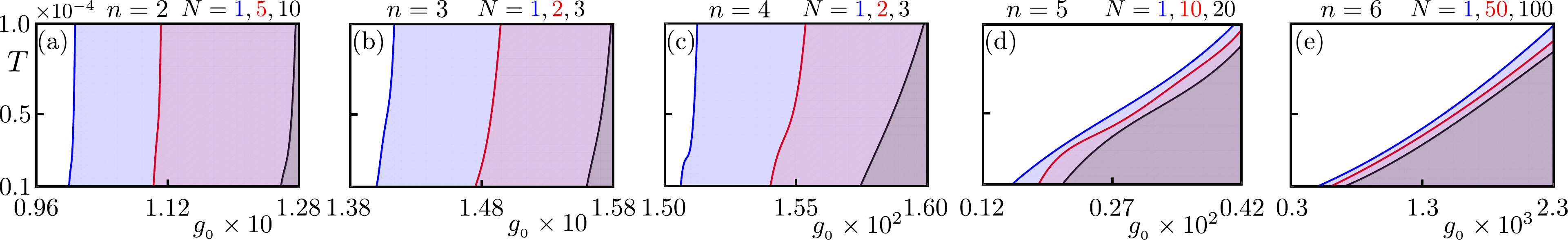}
    \caption{Cuts of the global phase diagram for the quarter metal in chirally-stacked $n$ layer graphene with $n \geq 2$ in the presence of a repulsive density-density interaction ($g_{_0}$) when the normal state (white shaded region) features annular Fermi rings at finite chemical potential ($\mu=2 \times 10^{-n}$) for (a) $n=2$, (b) $n=3$, (c) $n=4$, (d) $n=5$, and (e) $n=6$ with $\alpha_0= 10^{1-n}$ and $\alpha_3^n=10^{2n-2}$ (see text for notations). The vertical axis corresponds to temperature ($T$). Here all the quantities are dimensionless (see text for their definition). The colored regions represent a pair density wave for different flavor number $N$ (color coded), and its boundary with the normal state also marks the transition temperature ($T_c$) for the paired state.        
    }~\label{fig:PhaseDiag_Annular}
\end{figure*}

\begin{figure}[t!]
    \includegraphics[width=1.00\linewidth]{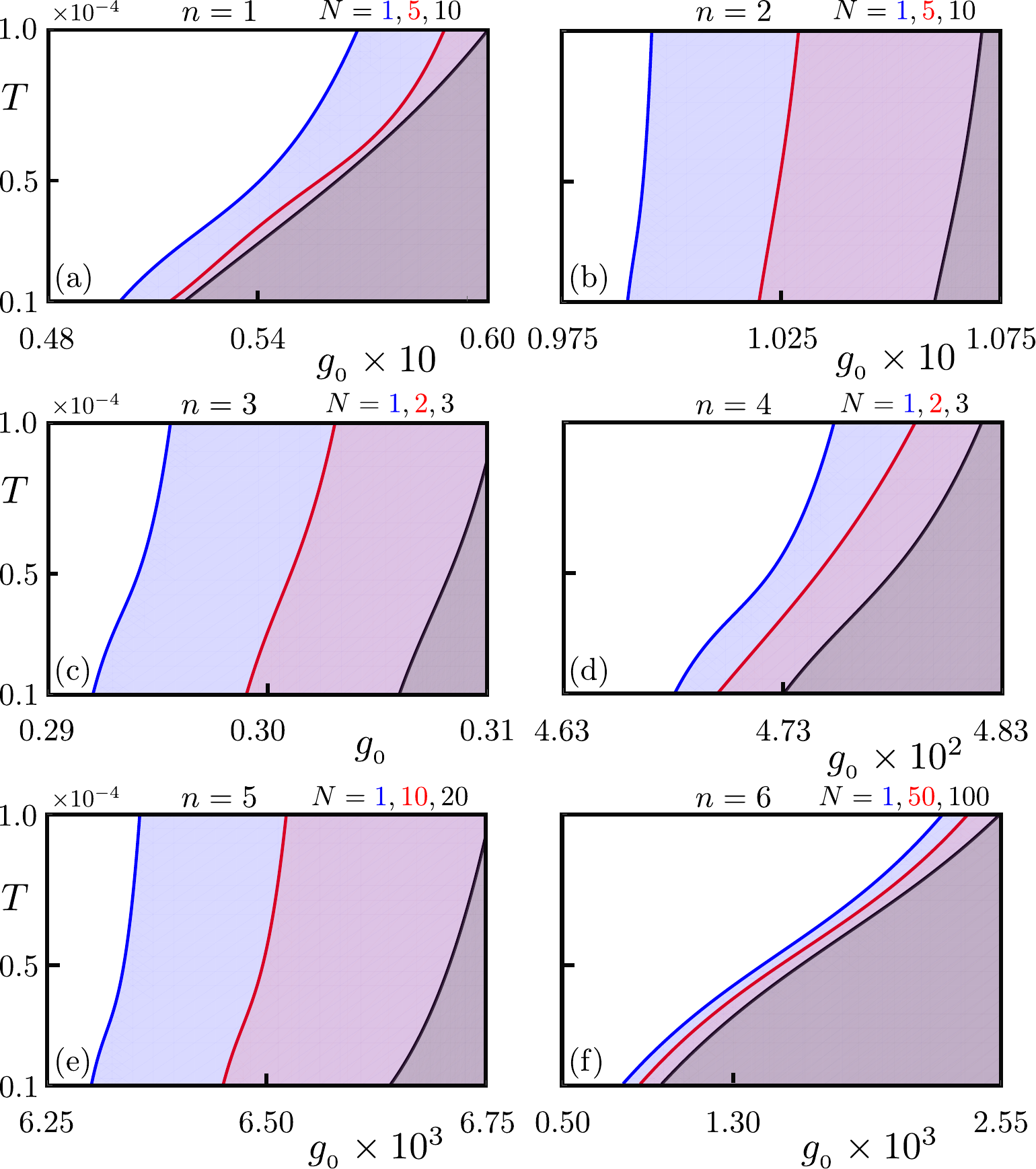}
    \caption{Same as Fig.~\ref{fig:PhaseDiag_Annular} but in the presence of a simply connected Fermi ring in the normal state. Here we show results for (a) $n=1$ (monolayer graphene) for which we set $\alpha_0= 0.03$ and $\mu=0.08$, and (b)-(f) $n=2$, $n=3$, $n=4$, $n=5$, and $n=6$, respectively, by setting $\alpha_0= 10^{1-n}$, $\alpha_3^n=10^{2n-2}$, and $\mu=2 \times 10^{1-n}$. Rest of the details are the same as in Fig.~\ref{fig:PhaseDiag_Annular}.  
    }~\label{fig:PhaseDiag_Simplyconnected}
\end{figure}

\emph{Results}.~In this work, from a leading-order renormalization group (RG) analysis, summarized in terms of the Feynman diagrams in Figs.~\ref{fig:Feynman_Interaction} and~\ref{fig:Feynman_Source}, we show that a repulsive local density-density interaction destabilizes the quarter metal toward the nucleation of such a PDW at low temperatures. This conclusion is insensitive to the underlying Fermi ring topology (annular or simply connected) in the normal state. While these findings are promising, shedding valuable insights into the possible microscopic origin of the observed superconductivity in the close proximity to the quarter metal, we emphasize that the purpose of the current investigation is not to predict nonuniversal quantities such as the superconducting transition temperature ($T_c$). Nonetheless, we construct cuts of the associated phase diagrams in the $(g_{_0}, T)$ plane with parameter values that are reasonably consistent with experimental data, where $g_{_0}$ ($T$) is the dimensionless coupling constants (temperature), defined shortly. The results are shown in Figs.~\ref{fig:PhaseDiag_Annular} and~\ref{fig:PhaseDiag_Simplyconnected}.

\emph{Background}.~Over the last few years, tremendous experimental breakthroughs have unfolded the effects of electronic interactions in the global phase diagram of chirally-stacked $n$ layer graphene heterostructures for $n \geq 2$ when these systems are subject to strong $D$ fields and are gradually doped away from the charge neutrality point. At sufficiently high doping, electronic bands display the full four-fold valley and spin degeneracy, yielding a metal. By contrast, as the chemical doping or carrier density is gradually reduced, the system enters into a half-metallic phase, where the electronic bands retain their valley degeneracy but lose the spin degeneracy. Upon further lowering the chemical doping or carrier density, the residual valley degeneracy gets lifted from the bands and the system enters into a fully polarized quarter metal phase. These features have been observed in all the systems with $n \geq 2$, yielding a unified global phase diagram. Typically, the valley-polarized quarter metal results from the anomalous Hall ordering, manifesting via nonzero anomalous Hall conductivity ($\sigma_{xy}$) and hysteresis in off-diagonal resistivity ($R_{xy}$). However, the quarter metal can also result from an unusual valley coherent charge density-wave order, as observed in systems with $n=3$ and $n=6$~\cite{ref:BBLG:1, ref:BBLG:2, ref:BBLG:3, ref:BBLG:4, ref:RTLG:1, ref:RTLG:2, ref:RTLG:3, ref:RTLG:4, ref:RTLG:5, ref:TTLG:1, ref:TTLG:2, ref:RPLG:1, ref:RPLG:2, ref:RPLG:3, ref:RHLG:2, ref:RHLG:3, ref:RHLG:4, ref:RHLG:5}. 

%

Irrespective of the nature of the orderings, leading to the formation of such fractional metals, they serve as the parent states, fostering various superconducting states. For example, in the close vicinity of the half-metal phase, a paired state devoid of the Pauli limiting magnetic field has been observed, suggesting their triplet nature. Such a paired state can arise from electron-phonon interactions or the Cooper glue can be mediated by incipient antiferromagnetic fluctuations, which can be responsible for the realization of the parent half-metal when it acquires a global expectation value, with the assistance from $D$ fields. More recently, superconducting states near the quarter metal have been observed in systems with $n=4$ and $n=6$. At least when the quarter metal is valley polarized, the fermionic degrees of freedom bear a momentum ${\bf K}$ and their Cooper pairs thus carry a finite momentum of $2 {\bf K}$, yielding a PDW. All these experimental observations triggered a surge of theoretical investigations exploring various aspects of the global phase diagram of such a family of systems~\cite{ref:RTLGnew:1, ref:RTLGnew:2, ref:RTLGnew:3, ref:RTLGnew:4, ref:RTLGnew:5, ref:RTLGnew:6, ref:RTLGnew:7, ref:RTLGnew:8, ref:RTLGnew:10, ref:RTLGnew:11, ref:RTLGnew:12, ref:RTLGnew:13, ref:RTLGnew:14, ref:RTLGnew:16, ref:RTLGnew:18, ref:RTLGnew:19, ref:BBLGRTLGnew:3, ref:BBLGnew:1, ref:BBLGnew:3, ref:BBLGnew:4, ref:BBLGnew:5, ref:BBLGnew:6, ref:BBLGnew:8, ref:BBLGnew:9, ref:BBLGRTLGTTLGnew:3, ref:BBLGRTLGTBLGnew:1, ref:MLGBBLGRTLGnew:1, ref:RTLGRHLGnew:1, ref:RNLGnew:1, ref:Thnew:1, ref:Thnew:2}. The present work aims to unfold a possible microscopic origin of the effective attractive interaction, responsible for the formation of Cooper pairs in the PDW channel.

\emph{Model}.~The effective low-energy model for the quarter metal in monolayer graphene accounts for the intralayer nearest-neighbor hopping ($t_0$), yielding gapless Dirac fermions~\cite{RMP:graphene}, and a staggered sublattice potential, producing a mass gap in the spectrum~\cite{semenoff}. For chirally-stacked $n$ layer graphene, besides $t_0$ we account for the interlayer dimer hopping ($t_\perp$) and external $D$ fields, entering the Hamiltonian as a potential difference between the layers. Out of $2n$ bands in such systems, $2n-2$ bands are separated by a band gap of $2 t_\perp$, which are integrated out and do not participate in the low-energy description. Two low-energy bands predominantly reside on the sites that are not connected by the dimer hopping.

We then arrive at the following universal effective two-band Hamiltonian for the spin and valley polarized quarter metal in the entire family of such systems $H_0=\tau_\nu d_\nu (\vec{k})$, where summation over repeated indices with $\nu=0, \cdots 3$ is assumed. Two-dimensional Pauli matrices $\{ \tau_\nu \}$ operate on the sublattice or layer index, $d_1(\vec{k})= \alpha_n |\vec{k}|^n \cos(n \phi_{\vec{k}})$, $d_2(\vec{k})= \alpha_n |\vec{k}|^n \sin(n \phi_{\vec{k}})$, $d_3(\vec{k})=V\left(1- \alpha_3^n |\vec{k}|^{2n-2}\right)$, $d_0(\vec{k})=\mu$ is the chemical potential measured from the charge neutrality point, $\alpha_n=\left( \sqrt{3}t_0/2\right)^n/t^{n-1}_\perp$, $\tan(\phi_{\vec{k}})=k_y/k_x$, $V$ is the uniform staggered potential between two sublattices for $n=1$ or the electrostatic potential difference between the top and bottom layers for $n \geq 2$ with $V=(n-1) D d$, where $d$ is the interlayer distance, and $\alpha^n_3=(1-\delta_{n,1}) (\sqrt{3} t_0/[2t_\perp])^{2n-2}$ with $\delta_{n,1}$ as the Kronecker delta symbol~\cite{ref:Thnew:2}. Hence, the quarter metal features annular (simply-connected) Fermi rings for small (large) $\mu$ for $n \geq 2$. The Fermi ring is always simply-connected for $n=1$. We set the lattice spacing $a=1$.

\emph{Interactions}.~We consider generic local four-fermion interactions, captured by the imaginary time ($\tau$) Euclidean action $S_{\rm int}=\int d^2\vec{r} d\tau \; L_{\rm int}$ with 
\begin{equation}
    L_{\rm int}= g_{_0} (\psi^\dagger \tau_{0}\psi)^2 
               + g_{_\perp} \sum_{j=1,2} (\psi^\dagger \tau_{j} \psi)^2 
               +g_{_3} (\psi^\dagger \tau_{3}\psi)^2,
\end{equation}
where $\psi$ and $\psi^\dagger$ are two-component spinors. However, due to the Fierz identity among the quartic terms, the number of linearly independent local four-fermion interactions is only \emph{one}, which we choose to be $(\psi^\dagger \tau_{0}\psi)^2$ without any loss of generality~\cite{Fierz:0, Fierz:1, Fierz:2}. Therefore, any microscopic interaction can be cast in terms of a suitable bare or initial condition of $g_{_0}$. Throughout, we take the \emph{bare} density-density interaction ($g_{_0}$) to be repulsive in nature without delving into its microscopic origin. As all the orderings (fractional metals and superconductivity) always take place at finite doping, where the density of states is finite, the conventional Thomas-Fermi screening converts any long-range interaction (such as Coulomb repulsion) into a local one, as considered here. Although, the bare local interactions are instantaneous, their frequency dependence develops during the coarse grain, which we account for through RG analysis, discussed next.

To capture the impact of such a repulsive interaction on the quarter metal, we integrate out fast Fourier modes living within the Wilsonian momentum shell $\Lambda e^{-\ell} < |\vec{k}| < \Lambda$, where $\Lambda$ is an ultraviolet momentum cutoff up to which the dispersion scales as $|\vec{k}|^n$ and $\ell$ is the logarithm of the RG scale. The corresponding Feynman diagrams are shown in Fig.~\ref{fig:Feynman_Interaction}. After performing the summation over the fermionic Matsubara frequencies, we arrive at the differential RG flow equation for the dimensionless coupling constant $g_{_0}\Lambda^{2-n}/(8 \pi \alpha_n) \to g_{_0}$, given by
\begin{equation}~\label{eq:RGflowcoupling}
\frac{dg_{_0}}{d\ell}= (n-2)g_{_0}+\sum_{\tau=\pm} \left[ \left(N-1\right) F_{\tau}(T,\mu)+  F^n_{\tau}( T,\mu)\right] g_{_0}^2.
\end{equation}
Here, we have incorporated a flavor number $N$ for two-component fermions to gain control over the RG calculation and quantum fluctuations about the mean-field or saddle point solutions, obtained by taking $N \to \infty$~\cite{RGscheme:0}. Various functions appearing in Eq.~\eqref{eq:RGflowcoupling}, are defined in terms of the dimensionless quantities $T/\left(\Lambda^n \alpha_n\right) \to T$ (temperature), $\mu/\left(\Lambda^n \alpha_n\right) \to \mu$, $V/\left(\Lambda^n \alpha_n\right) \to \alpha_0$, and $\alpha_3^n \Lambda^{2n-2} \to \alpha_3^n$ and the parameters $E_{\Lambda}=(1+ \alpha_0^2(1 - \alpha_3^n)^2)^{1/2} \equiv (1+\tilde{d_3}^2)^{1/2}$ and $E^\tau_\Lambda=E_\Lambda + \tau \mu$ as
\begin{widetext}
\begin{eqnarray}
&&F_{\tau} ( T,\mu) = \frac{1}{ T}\operatorname{sech}^2\left(\frac{E^\tau_{\Lambda}}{2 T}\right), \: 
G_\tau( T,\mu)=\operatorname{tanh}\left(\frac{E^\tau_{\Lambda}}{2 T}\right), \:\: \text{and} \:\: \nonumber \\
&&F_{\tau}^n ( T,\mu) = \frac{F_\tau( T,\mu)}{2} \left[\frac{1}{2E_{\Lambda}^2}+\frac{\tilde{d_3}^2}{E_{\Lambda}^2}-1\right] 
   +  G_\tau( T,\mu) \bigg[ 1 
   +    \tau \frac{E_{\Lambda} E^\tau_\Lambda- 2\mu^2}{  \mu \left( E_{\Lambda}^2-\mu^2\right)}
   +  \frac{E_{\Lambda}^2-\tau  E_{\Lambda} \mu- \mu^2}{ E_{\Lambda}^3 \mu \left( E_{\Lambda}^2-\mu^2\right)}\left\{  \tau \frac{\left(-1\right)^n}{2}+\tilde{d_3}^2\right\} \bigg].
\end{eqnarray}
\end{widetext}
.

In parallel, we also consider the RG flows of the dimensionless parameters that are given by 
\begin{equation}
\frac{d x }{ d \ell} = n x 
\:\:\: \text{and} \:\:\:
\frac{d \alpha_3^n }{ d \ell} = \left(2-2n\right) \alpha_3^n
\end{equation}
for $x= T,\mu,$ and $\alpha_0$, which we solve to find their running or $\ell$-dependent values, entering Eq.~\eqref{eq:RGflowcoupling}, for various initial conditions or bare values at $\ell=0$ (always set to be less than unity for the validity of the perturbative RG analysis). The RG flows for $x= T,\mu,$ and $\alpha_0$ yield infrared cutoffs $\ell^\star_x=-\ln [x(0)]/n$, which in turn determine the infrared (IR) cutoff for the RG flow of $g_{_0}(\ell)$, determined by $\ell^\star_{\rm IR}={\rm min}. (\ell^\star_t, \ell^\star_\mu, \ell^\star_{\alpha_0})$. When for a given set of initial or bare values $g(\ell^\star_{\rm IR})$ \emph{diverges} toward $\pm \infty$, that indicates a breakdown of the perturbative analysis with respect to the disordered phase, namely the quarter metal, and a concomitant onset of a broken symmetry phase. On the other hand, $g(\ell^\star)<1$ marks the stable normal state~\cite{RGscheme:1, RGscheme:2}. We follow this prescription to construct various cuts of the global phase diagram for the quarter metal of chirally stacked graphene multilayer in the $(g_{_0}, T)$ plane, shown in Fig.~\ref{fig:PhaseDiag_Annular} and Fig.~\ref{fig:PhaseDiag_Simplyconnected} with underlying annular and simply-connected Fermi rings, respectively.

Before delving into the nature of the ordered phase, it is worth highlighting a generic feature of the phase boundary with the quarter metal. Irrespective of $n$ and the geometry of Fermi rings, we note that such a phase boundary gets pushed toward stronger coupling with increasing $N$ (flavor number). Therefore, the onset of any ordered state is triggered by quantum fluctuations about the saddle point or mean-field state, which always describes the disordered phase. In this respect, the nucleation of the ordered phase is purely fluctuation-mediated. It should be noted that superconductivity is found for any finite $N$, including the physically pertinent scenario of $N=1$ (see Figs.~\ref{fig:PhaseDiag_Annular} and~\ref{fig:PhaseDiag_Simplyconnected}) and thus the flavor $N$ serves as a control parameter to account for fluctuations, following the general spirit of large-$N$ expansion~\cite{largeNReview}.

\emph{Pair density wave}.~To identify the nature of the ordered phase, we simultaneously run the RG flow equations for all the symmetry-allowed source terms, capturing various fermion bilinears. The corresponding imaginary time action for the particle-hole or excitonic orders reads as $S^{ph}_{s}=\int d^2 \vec{r} d\tau \; H^{ph}_{s}$, where
\begin{equation}
H^{ph}_{s} = \Delta_0 \left( \psi^\dagger \tau_0 \psi \right) + \Delta_\perp \sum^{2}_{j=1} \left( \psi^\dagger \tau_j \psi \right) + \Delta_3 \left( \psi^\dagger \tau_3 \psi \right). 
\end{equation}
The leading order renormalization of the source terms arises from the Feynman diagrams shown in Fig.~\ref{fig:Feynman_Source}. Notice that the term proportional to $\Delta_0$ couples to fermionic density, which does not break any microscopic symmetry and its renormalization gives a renormalized $\mu$, entering Eq.~\eqref{eq:RGflowcoupling}. Since such a renormalization of $\mu$ is proportional to $g_{_0}$, we neglect it in the spirit of the leading order RG analysis. For the same reason, we also neglect the contribution from the Feynman diagram in Fig.~\ref{fig:Feynman_Interaction}(b). The leading order differential RG flow equations for the remaining two source terms are   
\begin{eqnarray}
\bar{\beta}_{\Delta_\perp} &=& -\sum_{\tau=\pm} \left[ \left( 1 +\frac{\tilde{d_3}^2}{E_{\Lambda}^2}\right) F_\tau( T,\mu) + \frac{1}{E_{\Lambda}^3} \; G_\tau( T,\mu) \right] g_{_0} \nonumber \\
\text{and} \: \bar{\beta}_{\Delta_3} &=& -\frac{1}{2 E_\Lambda} \sum_{\tau=\pm} G_\tau(T, \mu) g_{_0},
\end{eqnarray}
where $\bar{\beta}_y \equiv d\ln(y)/d\ell-n$. To account for the possible superconductivity or particle-particle ordering, we also consider the corresponding imaginary time action $S^{pp}_{s}=\int d^2 \vec{r} d\tau (\Delta_p \psi^\dagger \tau_2 \psi^\star + H.c.)$. The leading order differential RG flow equation for $\Delta_p$ takes the form
\begin{eqnarray}
\bar{\beta}_{\Delta_p} &=& \frac{g_{_0}}{2} \sum_{\tau=\pm} \bigg[ \frac{G_\tau( T,\mu)}{\mu} \bigg\{ \frac{E^\tau_\Lambda \left[\left(-1\right)^{n+1}-\tilde{d_3}^2\right]}{ E_{\Lambda} \left( E_{\Lambda}^2-\mu^2\right)} \nonumber \\ 
&+& \frac{E_{\Lambda} E^\tau_\Lambda - 2\mu^2}{\left( E_{\Lambda}^2-\mu^2\right)} \bigg\} \bigg].  
\end{eqnarray}

We run the RG flow equations for $g_{_0}$ for various initial conditions at $\ell=0$, and simultaneously track the flow of  $\Delta_\perp$, $\Delta_3$, and $\Delta_p$. When $g_{_0}(\ell^\star) \to \pm \infty$, the nature of the ordered phase is determined by the source term that diverges toward $+\infty$ in the \emph{fastest} fashion, thereby acquiring the \emph{largest} scaling dimension. We note that for bare repulsive interaction in the presence of a Fermi ring (annular or simply-connected), an ordered phase develops only beyond critical interaction strengths, which, however, always corresponds to the superconducting phase that is PDW in nature. We note that the phase boundary between a quarter metal and the PDW moves toward stronger coupling with increasing $N$, indicating that the pairing is solely triggered by quantum fluctuations. It should be noted that the formation of PDW from repulsive interaction follows the spirit of the Kohn-Luttinger mechanism~\cite{kohnluttinger}, which is typically studied in normal single-band systems in the absence of any interband scattering, devoid of any topological ingredient, such as the chirality of order $n$ as is the case in the systems, we consider here. Finally, it should be noted that due to the reduced dimensionality of these systems, the transition temperature predicted from the RG analysis marks only a crossover temperature, where the Cooper pairs form. However, their coherent nucleation takes place at a lower temperature via the Kosterlitz-Thouless transition through vortex-antivortex binding~\cite{KT}.  

\emph{Real systems}.~Here, we consider a minimal Hamiltonian for the quarter metal in a chirally-stacked $n$ layer graphene, which accounts for the dispersion that scales as $|\vec{k}|^n$, resulting from the dominant intralayer NN and interlayer dimer hopping processes, and the imprints of $D$ fields, responsible for distinct Fermi ring geometry therein. However, such models neglect various sub-dominant hopping elements, which, for example, yield trigonal warping, typically captured by terms that scale as $|\vec{k}|^m$ with $m<n$. This raises a question regarding the relevance of our findings in real graphene-based multilayer systems, where superconductivity near the quarter metal has been observed in experiments~\cite{PDWExp:1, PDWExp:2, PDWExp:3}.

Carrying a RG analysis with all powers of dispersion in the Green's function is a daunting task, which is somewhat unnecessary. Rather, we put forward a qualitative argument based on the results discussed so far to endorse the validity of our findings in realistic situations. Notice that different powers of momentum dominate over different ranges of $|\vec{k}|$. Therefore, we can split the entire momentum integrals into different segments, over each of which only one power of $|\vec{k}|$ dominates and the corresponding Hamiltonian ($H_0$) or Green's function enjoys a full rotational symmetry~\cite{ref:Thnew:2}. As an example, let us assume that dispersion proportional to $|\vec{k}|^p$ ($|\vec{k}|^q$) dominates for $|\vec{k}|>k_{pq}$ ($|\vec{k}|<k_{pq}$). Then we perform the RG analysis with only $|\vec{k}|^p$ ($|\vec{k}|^q$) dispersion over the region $|\vec{k}|>k_{pq}$ ($|\vec{k}|<k_{pq}$). The running coupling $g_{_0} (\ell)$ at the scale $|\vec{k}|=k_{pq}$, obtained from the RG flow of $g_{_0}$ with only $|\vec{k}|^p$ dispersion, serves as the initial condition for the RG flow of $g_{_0}$ for $|\vec{k}|<k_{pq}$. Given that, irrespective of the power of $|\vec{k}|$ in the dispersion, the density-density repulsion yields superconductivity in the PDW channel, one can conclude that such an outcome remains operative even when the full dispersion, containing various powers of momentum, is taken into account.

\emph{Summary and discussions}.~To summarize, from a leading order RG analysis, controlled by the flavor number ($N$), here we show that repulsive density-density interaction among spin and valley polarized fermionic excitations can destabilize a quarter metal toward the nucleation of a unique local superconducting order that represents a PDW with a characteristic $2 {\bf K}$ periodicity. We show that the effective attractive interaction responsible for the Cooper pair condensation is solely driven by quantum fluctuations about the stable mean-field or saddle point solution that describes the parent normal state (quarter metal). Such conclusions are insensitive to the Fermi ring topology (annular or simply-connected). We also argue that our conclusions from simple two-band models with the leading-order in momentum dispersion can rationally be extended to shed light on the possible microscopic origin of the pairing interaction in quarter metal of chirally-stacked $n$ layer graphene, responsible for the experimentally observed superconductivity in its close proximity in tetralayer and hexalayer graphene, in a more realistic situation involving distant hopping elements, yielding trigonal warping, for example. We are optimistic that with improved sample quality the PDW can be observed in other members of this family.

Optical monolayer graphene of neutral atoms stands as the most promising platform where similar physics can be observed. The requisite valley and spin degeneracy lifting therein can be accomplished with Hubbard repulsion mediated antiferromagnetic ordering~\cite{optgra:1} in the presence of externally engineered small staggered potential difference between its two sublattices~\cite{optgra:2} and Haldane's anomalous Hall order~\cite{haldane, optgra:3} as these orders mutually \emph{commute}~\cite{ref:RTLGnew:10, ref:BBLGnew:3}. Present discussion should stimulate future experiments on such a platform, gearing toward harnessing a PDW superfluid on monolayer graphene.

\emph{Acknowledgments}.~This work was supported by NSF CAREER Grant No. DMR-2238679 of B.R. We thank Christopher A.\ Leong for comments on the manuscript.  

\emph{Data availability}.~The numerical codes and data that support the findings of this article are openly
available in Ref.~\cite{dataavailability}.


\begin{thebibliography}{}

\bibitem{PDWgraphene:1} S.\ A.\ Murshed, B.\ Roy, Nodal pair density waves from a quarter-metal in crystalline graphene multilayers, \href{https://journals.aps.org/prb/abstract/10.1103/wy3f-hgr9}{Phys.\ Rev.\ B \textbf{112}, 085121 (2025)}.

\bibitem{PDWgraphene:2} B.\ Roy and I.\ F.\ Herbut, Unconventional superconductivity on honeycomb lattice: Theory of Kekule order parameter, \href{https://journals.aps.org/prb/abstract/10.1103/PhysRevB.82.035429}{Phys.\ Rev.\ B {\bf 82}, 035429 (2010)}.


\bibitem{PDWExp:1} E.\ Morissette, P.\ Qin, H.-T.\ Wu, N.\ J.\ Zhang, R.\ Q.\ Nguyen, K.\ Watanabe, T.\ Taniguchi, and J.\ I.\ A.\ Li, Striped Superconductor in Rhombohedral Hexalayer Graphene, \href{https://arxiv.org/abs/2504.05129}{arXiv:2504.05129 (2025)}.

\bibitem{PDWExp:2} T.\ Han, Z.\ Lu, Z.\ Hadjri, L.\ Shi, Z.\ Wu, W.\ Xu, Y.\ Yao, A.\ A.\ Cotten, O.\ S.\ Sedeh, H.\ Weldeyesus, J.\ Yang, J.\ Seo, S.\ Ye, M.\ Zhou, H.\ Liu, G.\ Shi, Z.\ Hua, K.\ Watanabe, T.\ Taniguchi, P.\ Xiong, D.\ M.\ Zumbühl, L.\ Fu, and L.\ Ju, Signatures of Chiral Superconductivity in Rhombohedral Graphene, \href{https://www.nature.com/articles/s41586-025-09169-7}{Nature \textbf{643}, 654 (2025)}.

\bibitem{PDWExp:3} Y.\ Choi, Y.\ Choi, M.\ Valentini, C.\ L.\ Patterson, L.\ F.\ W.\ Holleis, O.\ I.\ Sheekey, H.\ Stoyanov, X.\ Cheng, T.\ Taniguchi, K.\ Watanabe and A.\ F.\ Young, Superconductivity and quantized anomalous Hall effect in rhombohedral graphene, \href{https://www.nature.com/articles/s41586-025-08621-y}{Nature \textbf{639}, 342 (2025)}.



\bibitem{ref:TTLGnew:1}  Y.\-Z.\ Chou, J.\ Zhu, and S.\ Das Sarma, Intravalley spin-polarized superconductivity in rhombohedral tetralayer graphene, \href{https://journals.aps.org/prb/abstract/10.1103/PhysRevB.111.174523}{Phys.\ Rev.\ B \textbf{111}, 174523 (2025)}.

\bibitem{ref:TTLGnew:2} A.\ S.\ Patri and M.\ Franz, Family of multilayer graphene superconductors with tunable chirality: Momentum-space vortices nucleated by a ring of Berry curvature, \href{https://journals.aps.org/prb/abstract/10.1103/pgqh-wm5v}{Phys.\ Rev.\ B {\bf 112}, 214505 (2025)}.

\bibitem{ref:TTLGnew:3} M.\ Christos, P.\ M.\ Bonetti, M.\ S.\ Scheurer, Finite-momentum pairing and superlattice superconductivity in valley-imbalanced rhombohedral graphene, \href{https://arxiv.org/abs/2503.15471}{arXiv:2503.15471 (2025)}.

\bibitem{ref:TTLGnew:4} Z.\ Dong, P.\ A.\ Lee, A controlled expansion for pairing in a polarized band with strong repulsion, \href{https://arxiv.org/abs/2503.11079}{arXiv:2503.11079 (2025)}.

\bibitem{ref:TTLGnew:5} G.\ Parra-Mart\'inez, A.\ Jimeno-Pozo, V.\ T.\ Phong, H.\ Sainz-Cruz, D.\ Kaplan, P.\ Emanuel, Y.\ Oreg, P.\ A.\ Pantale\'on, J.\ \'A.\ Silva-Guill\'en, and F.\ Guinea, Band Renormalization, Quarter Metals, and Chiral Superconductivity in Rhombohedral Tetralayer Graphene, \href{https://journals.aps.org/prl/abstract/10.1103/zfmh-rjzc}{Phys.\ Rev.\ Lett.\ \textbf{135}, 136503 (2025)}.

\bibitem{ref:TTLGnew:6} Z.\ Han, J.\ Herzog-Arbeitman, Q.\ Gao, and E.\ Khalaf, Exact models of chiral flat-band superconductors, \href{https://arxiv.org/abs/2508.21127}{arXiv:2508.21127 (2025)}.

\bibitem{ref:TTLGnew:7} Y.\ Barlas, F.\ Zhang, and E.\ Rossi, Quantum Geometry Induced Kekulé Superconductivity in Haldane phases, \href{https://arxiv.org/abs/2508.21791v1}{arXiv:2508.2179 (2025)}.

\bibitem{ref:TTLGnew:8} H.\ Yang and Y-H.\ Zhang, Topological incommensurate Fulde-Ferrell-Larkin-Ovchinnikov superconductor and Bogoliubov Fermi surface in rhombohedral tetralayer graphene, \href{https://journals.aps.org/prb/abstract/10.1103/k8s3-dgfs}{Phys.\ Rev.\ B {\bf 112}, L020506 (2025)}.

\bibitem{ref:TTLGnew:9} M.\ Geier, M.\ Davydova, and L.\ Fu, Chiral and topological superconductivity in isospin polarized multilayer graphene, \href{https://www.nature.com/articles/s41467-025-66902-6}{Nat.\ Commun.\ {\bf 17}, 232 (2026).}.







\bibitem{ref:BBLG:1} H.\ Zhou, L.\ Holleis, Y.\ Saito, L.\ Cohen, W.\ Huynh, C.\ L.\ Patterson, F.\ Yang, T.\ Taniguchi, K.\ Watanabe, and A.\ F.\ Young, Isospin magnetism and spin-polarized superconductivity in Bernal bilayer graphene, \href{https://www.science.org/doi/10.1126/science.abm8386}{Science \textbf{375}, 774 (2022)}.

\bibitem{ref:BBLG:2} S.\ C.\ de la Barrera, S.\ Aronson, Z.\ Zheng, K.\ Watanabe, T.\ Taniguchi, Q.\ Ma, P.\ Jarillo-Herrero, and R.\ Ashoori, Cascade of isospin phase transitions in Bernal-stacked bilayer graphene at zero magnetic field, infrared spectroscopy, \href{https://www.nature.com/articles/s41567-022-01616-w}{Nat.\ Phys.\ \textbf{18}, 771 (2022)}.

\bibitem{ref:BBLG:3} A.\ M.\ Seiler, F.\ R.\ Geisenhof, F.\ Winterer, K.\ Watanabe, T.\ Taniguchi, T.\ Xu, F.\ Zhang, and R.\ T.\ Weitz, Quantum cascade of correlated phases in trigonally warped bilayer graphene,
\href{https://www.nature.com/articles/s41586-022-04937-1}{Nature (London) \textbf{608}, 298 (2022)}.

\bibitem{ref:BBLG:4} C.\ Li, F.\ Xu, B.\ Li, J.\ Li, G.\ Li, K.\ Watanabe, T.\ Taniguchi, B.\ Tong, J.\ Shen, L.\ Lu, J.\ Jia, F.\ Wu, X.\ Liu and T.\ Li, Tunable superconductivity in electron- and hole-doped Bernal bilayer graphene, \href{https://www.nature.com/articles/s41586-024-07584-w}{Nature \textbf{631}, 300 (2024)}.



\bibitem{ref:RTLG:1} H.\ Zhou, T.\ Xie, A.\ Ghazaryan, T.\ Holder, J.\ R.\ Ehrets, E.\ M.\ Spanton, T.\ Taniguchi, K.\ Watanabe, E.\ Berg, M.\ Serbyn, and A.\ F.\ Young, Half and quarter metals in rhombohedral trilayer graphene, \href{https://www.nature.com/articles/s41586-021-03938-w}{Nature (London) \textbf{598}, 429 (2021)}.

\bibitem{ref:RTLG:2} H.\ Zhou, T.\ Xie, T.\ Taniguchi, K.\ Watanabe, and A.\ F.\ Young, Superconductivity in rhombohedral trilayer graphene, \href{https://www.nature.com/articles/s41586-021-03926-0}{Nature (London) \textbf{598}, 434 (2021)}.

\bibitem{ref:RTLG:3} C.\ L.\ Patterson, O.\ I.\ Sheekey, T.\ B.\ Arp, L.\ F.\ W.\ Holleis, J.\ Koh, Y.\ Choi, T.\ Xie, S.\ Xu, Y.\ Guo, H.\ Stoyanov, E.\ Redekop, C.\ Zhang, G.\ Babikyan, D.\ Gong, H.\ Zhou, X.\ Cheng, T.\ Taniguchi, K.\ Watanabe, M.\ E.\ Huber, C.\ Jin, É.\ L.\-Hurtubise, J.\ Alicea and A.\ F.\ Young, Superconductivity and spin canting in spin–orbit-coupled trilayer graphene, \href{https://www.nature.com/articles/s41586-025-08863-w}{Nature (London) \textbf{641}, 632 (2025)}.

\bibitem{ref:RTLG:4} T.\ Arp, O.\ Sheekey, H.\ Zhou, C.\ L.\ Tschirhart, C.\ L.\ Patterson, H.\ M.\ Yoo, L.\ Holleis, E.\ Redekop, G.\ Babikyan, T.\ Xie, J.\ Xiao, Y.\ Vituri, T.\ Holder, T.\ Taniguchi, K.\ Watanabe, M.\ E.\ Huber, E.\ Berg and A.\ F.\ Young, Intervalley coherence and intrinsic spin–orbit coupling in rhombohedral trilayer graphene, \href{https://www.nature.com/articles/s41567-024-02560-7}{Nat.\ Phys.\ {\bf 20}, 1413 (2024)}.

\bibitem{ref:RTLG:5} Y.\ Liu, A.\ Gupta, Y.\ Choi, Y.\ Vituri, H.\ Stoyanov, J.\ Xiao, Y.\ Wang, H.\ Zhou, B.\ Barick, T.\ Taniguchi, K.\ Watanabe, B.\ Yan, E.\ Berg, A.\ F.\ Young, H.\ Beidenkopf, N.\ Avraham, Visualizing incommensurate inter-valley coherent states in rhombohedral trilayer graphene, \href{https://arxiv.org/abs/2411.11163}{arXiv: 2411.11163 (2024)}.


 
\bibitem{ref:TTLG:1} K.\ Liu, J.\ Zheng, Y.\ Sha, B.\ Lyu, F.\ Li, Y.\ Park, Y.\ Ren, K.\ Watanabe, T.\ Taniguchi, J.\ Jia, W.\ Luo, Z.\ Shi, J.\ Jung and G.\ Chen, Spontaneous broken-symmetry insulator and metals in tetralayer rhombohedral graphene, \href{https://www.nature.com/articles/s41565-023-01558-1}{Nat.\ Nanotechnol.\ \textbf{19}, 188 (2024)}.

\bibitem{ref:TTLG:2} Y.\ Sha, J.\ Zheng, K.\ Liu, H.\ Du, K.\ Watanabe, T.\ Taniguchi, J.\ Jia, Z.\ Shi, R.\ Zhong, and G.\ Chen, Observation of a Chern insulator in crystalline ABCA-tetralayer graphene with spin-orbit coupling, \href{https://www.science.org/doi/full/10.1126/science.adj8272}{Science \textbf{384}, 414 (2024)}.



\bibitem{ref:RPLG:1} T.\ Han, Z.\ Lu, G.\ Scuri, J.\ Sung, J.\ Wang, T.\ Han, K.\ Watanabe, T.\ Taniguchi, H.\ Park, and L.\ Ju, Correlated insulator and Chern insulators in pentalayer rhombohedral-stacked graphene, \href{https://www.nature.com/articles/s41565-023-01520-1}{Nat.\ Nanotechnol.\ \textbf{19}, 181 (2024)}.

\bibitem{ref:RPLG:2} Z.\ Lu, T.\ Han, Y.\ Yao, A.\ P.\ Reddy, J.\ Yang, J.\ Seo, K.\ Watanabe, T.\ Taniguchi, L.\ Fu and L.\ Ju, Fractional quantum anomalous Hall effect in multilayer graphene, \href{https://www.nature.com/articles/s41586-023-07010-7}{Nature \textbf{626}, 759 (2024)}.

\bibitem{ref:RPLG:3} T.\ Han, Z.\ Lu, G.\ Scuri, J.\ Sung, J.\ Wang, T.\ Han, K.\ Watanabe, T.\ Taniguchi, L.\ Fu, H.\ Park and L.\ Ju, Orbital multiferroicity in pentalayer rhombohedral graphene, \href{https://www.nature.com/articles/s41586-023-06572-w}{Nature \textbf{623}, 41 (2023)}.





\bibitem{ref:RHLG:2} E.\ Morissette, P.\ Qin, K.\ Watanabe, T.\ Taniguchi, and J.\ I.\ A.\ Li, Evidence of momentum space condensation in rhombohedral hexalayer graphene, \href{https://arxiv.org/abs/2503.09954}{arXiv:2503.09954 (2025)}.

\bibitem{ref:RHLG:3} J.\ Deng, J.\ Xie, H.\ Li, T.\ Taniguchi, K.\ Watanabe, J.\ Shan, K.\ F.\ Mak, and X.\ Liu, Superconductivity and Ferroelectric Orbital Magnetism in Semimetallic Rhombohedral Hexalayer Graphene, \href{https://arxiv.org/abs/2508.15909}{arXiv:2508.15909 (2025)}.

\bibitem{ref:RHLG:4} K.\ Kr\"otzsch, A.\ Herasymchuk, Y.\ Zhumagulov, A.\ Magrez, K.\ Watanabe, T.\ Taniguchi, S.\ G.\ Sharapov, O.\ V.\ Yazyev, and M.\ Banerjee, Magnetoelectric Switching of Competing Magnetic Orders in Rhombohedral Graphene, \href{https://arxiv.org/abs/2509.24672}{arXiv:2509.24672 (2025)}.

\bibitem{ref:RHLG:5} R.\ Q.\ Nguyen, H.-T.\ Wu, E.\ Morissette, N.\ J.\ Zhang, P.\ Qin, K.\ Watanabe, T.\ Taniguchi, A.\ W.\ Hui, D.\ E.\ Feldman and J.\ I.\ A.\ Li, A Hierarchy of Superconductivity and Topological Charge Density Wave States in Rhombohedral Graphene, \href{https://arxiv.org/abs/2507.22026}{arXiv:2507.22026 (2025)}.


\bibitem{ref:RTLGnew:1} Y.-Z.\ Chou, F.\ Wu, J.\ D.\ Sau, and S.\ Das Sarma, Acoustic-phonon-mediated superconductivity in rhombohedral trilayer graphene, \href{https://journals.aps.org/prl/abstract/10.1103/PhysRevLett.127.187001}{Phys.\ Rev.\ Lett.\ \textbf{127}, 187001 (2021)}.

\bibitem{ref:RTLGnew:2} S.\ Chatterjee, T.\ Wang, E.\ Berg, and M.\ P.\ Zaletel, Inter-valley coherent order and isospin fluctuation mediated superconductivity in rhombohedral trilayer graphene, \href{https://www.nature.com/articles/s41467-022-33561-w}{Nat.\ Commun.\ \textbf{13}, 6013 (2022)}.

\bibitem{ref:RTLGnew:3}  A.\ Ghazaryan, T.\ Holder, M.\ Serbyn, and E.\ Berg, Unconventional superconductivity in systems with annular fermi surfaces: Application to rhombohedral trilayer graphene, \href{https://journals.aps.org/prl/abstract/10.1103/PhysRevLett.127.247001}{Phys.\ Rev.\ Lett.\ \textbf{127}, 247001 (2021)}.

\bibitem{ref:RTLGnew:4} Y.\ Vituri, J.\ Xiao1, K.\ Pareek, T.\ Holder, and E.\ Berg, Incommensurate intervalley coherent states in ABC graphene: Collective modes and superconductivity, \href{https://journals.aps.org/prb/abstract/10.1103/PhysRevB.111.075103}{Phys.\ Rev.\ B \textbf{111}, 075103 (2025)}.

\bibitem{ref:RTLGnew:5} Y.\ Zhumagulov, D.\ Kochan, and J.\ Fabian, Emergent Correlated Phases in Rhombohedral Trilayer Graphene Induced by Proximity Spin-Orbit and Exchange Coupling, \href{https://journals.aps.org/prl/abstract/10.1103/PhysRevLett.132.186401}{Phys.\ Rev.\ Lett.\ \textbf{132}, 186401 (2024)}.

\bibitem{ref:RTLGnew:6} W.\ Qin, C.\ Huang, T.\ Wolf, N.\ Wei, I.\ Blinov, and A.\ H.\ MacDonald, Functional Renormalization Group Study of Superconductivity in Rhombohedral Trilayer Graphene, \href{https://journals.aps.org/prl/abstract/10.1103/PhysRevLett.130.146001}{Phys.\ Rev.\ Lett.\ \textbf{130}, 146001 (2023)}.

\bibitem{ref:RTLGnew:7} D.-C.\ Lu, T.\ Wang, S.\ Chatterjee, and Y.\!-Z.\ You, Correlated metals and unconventional superconductivity in rhombohedral trilayer graphene: A renormalization group analysis, \href{https://journals.aps.org/prb/abstract/10.1103/PhysRevB.106.155115}{Phys.\ Rev.\ B \textbf{106}, 155115 (2022)}.

\bibitem{ref:RTLGnew:8} Y.\ You and Ashvin Vishwanath, Kohn-Luttinger superconductivity and intervalley coherence in rhombohedral trilayer graphene, \href{https://journals.aps.org/prb/abstract/10.1103/PhysRevB.105.134524}{Phys.\ Rev.\ B \textbf{105}, 134524 (2022)}.

\bibitem{ref:RTLGnew:10} A.\ L.\ Szab\'o and B.\ Roy, Metals, fractional metals, and superconductivity in rhombohedral trilayer graphene, \href{https://journals.aps.org/prb/abstract/10.1103/PhysRevB.105.L081407}{Phys.\ Rev.\ B \textbf{105}, L081407 (2022)}.

\bibitem{ref:RTLGnew:11} T.\ Cea, P.\ A.\ Pantale\'on, V.\ T.\ Phong, and F.\ Guinea, Superconductivity from repulsive interactions in rhombohedral trilayer graphene: A Kohn-Luttinger-like mechanism, \href{https://journals.aps.org/prb/abstract/10.1103/PhysRevB.105.075432}{Phys.\ Rev.\ B \textbf{105}, 075432 (2022)}.

\bibitem{ref:RTLGnew:12} C.\ Huang, T.\ M.\ R.\ Wolf, W.\ Qin, N.\ Wei, I.\ V.\ Blinov, and A.\ H.\ MacDonald, Spin and orbital metallic magnetism in rhombohedral trilayer graphene, \href{https://journals.aps.org/prb/abstract/10.1103/PhysRevB.107.L121405}{Phys.\ Rev.\ B \textbf{107}, L121405 (2023)}.

\bibitem{ref:RTLGnew:13} V.\ Juri\v{c}i\'c, E.\ Mu\~noz, and R. Soto-Garrido, Optical conductivity as a probe of the interaction-driven metal in rhombohedral trilayer graphene, \href{https://www.mdpi.com/2079-4991/12/21/3727}{Nanomaterials \textbf{12}, 3727 (2022)}.

\bibitem{ref:RTLGnew:14} A.\ S.\ Patri and T.\ Senthil, Strong correlations in ABC-stacked trilayer graphene: Moir\'e is important, \href{https://journals.aps.org/prb/abstract/10.1103/PhysRevB.107.165122}{Phys.\ Rev.\ B \textbf{107}, 165122 (2023)}.

\bibitem{ref:RTLGnew:16} H.\ Dai, R.\ Ma, X.\ Zhang, T.\ Guo, and T.\ Ma, Quantum Monte Carlo study of superconductivity in rhombohedral trilayer graphene under an electric field, \href{https://journals.aps.org/prb/abstract/10.1103/PhysRevB.107.245106}{Phys.\ Rev.\ B \textbf{107}, 245106 (2023)}. 

\bibitem{ref:RTLGnew:18} C.\ W.\ Chau, S.\ A.\ Chen, and K.\ T.\ Law, Superconductivity from quasiparticle pairing of intervalley coherent state in rhombohedral trilayer graphene, \href{https://arxiv.org/abs/2404.19237}{arXiv:2404.19237 (2024)}.

\bibitem{ref:RTLGnew:19} Q.\-C.\ Zhang and J.\ Wang, Fermion-fermion interaction driven phase transitions in rhombohedral trilayer graphene, \href{https://journals.aps.org/prb/abstract/10.1103/PhysRevB.111.014111}{Phys.\ Rev.\ B \textbf{111}, 014111 (2025)}.


\bibitem{ref:BBLGRTLGnew:3} A.\ Jimeno-Pozo, H.\ Sainz-Cruz, T.\ Cea, P.\ A.\ Pantale\'on, and F.\ Guinea, Superconductivity from electronic interactions and spin-orbit enhancement in bilayer and trilayer graphene, \href{https://journals.aps.org/prb/abstract/10.1103/PhysRevB.107.L161106}{Phys.\ Rev.\ B \textbf{107}, L161106 (2023)}.


\bibitem{ref:BBLGnew:1} Y.-Z.\ Chou, F.\ Wu, and S.\ Das Sarma, Enhanced superconductivity through virtual tunneling in Bernal bilayer graphene coupled to $\rm WSe_2$, \href{https://journals.aps.org/prb/abstract/10.1103/PhysRevB.106.L180502}{Phys.\ Rev.\ B \textbf{106}, L180502 (2022)}.

\bibitem{ref:BBLGnew:3} A.\ L.\ Szab\'o and B.\ Roy, Competing orders and cascade of degeneracy lifting in doped Bernal bilayer graphene, \href{https://journals.aps.org/prb/abstract/10.1103/PhysRevB.105.L201107}{Phys.\ Rev.\ B \textbf{105}, L201107 (2022)}.

\bibitem{ref:BBLGnew:4} G.\ Wagner, Y.\ H.\ Kwan, N.\ Bultinck, S.\ H.\ Simon, and S.\ A.\ Parameswaran, Superconductivity from repulsive interactions in Bernal-stacked bilayer graphene, \href{https://journals.aps.org/prb/abstract/10.1103/PhysRevB.110.214517}{Phys.\ Rev.\ B \textbf{110}, 214517 (2024)}.

\bibitem{ref:BBLGnew:5} J.\ B.\ Curtis, N.\ R.\ Poniatowski, Y.\ Xie, A.\ Yacoby, E.\ Demler, and P.\ Narang, Stabilizing fluctuating spin-triplet superconductivity in graphene via induced spin-orbit coupling, \href{https://journals.aps.org/prl/abstract/10.1103/PhysRevLett.130.196001}{Phys.\ Rev.\ Lett.\ \textbf{130}, 196001 (2023)}.

\bibitem{ref:BBLGnew:6} Y.-Z.\ Chou, F.\ Wu, J.\ D.\ Sau, and S.\ Das Sarma, Acoustic-phonon-mediated superconductivity in Bernal bilayer graphene, \href{https://journals.aps.org/prb/abstract/10.1103/PhysRevB.105.L100503}{Phys.\ Rev.\ B \textbf{105}, L100503 (2022)}.

\bibitem{ref:BBLGnew:8} Y.\ Zhumagulov, D.\ Kochan, and J.\ Fabian, Swapping exchange and spin-orbit induced correlated phases in proximitized Bernal bilayer graphene, \href{https://journals.aps.org/prb/abstract/10.1103/PhysRevB.110.045427}{Phys.\ Rev.\ B \textbf{110}, 045427 (2024)}.

\bibitem{ref:BBLGnew:9} A.\ Fischer, L.\ Klebl, D.\ M.\ Kennes, and T.\ O.\ Wehling, Supercell Wannier functions and a faithful low-energy model for Bernal bilayer graphene, \href{https://journals.aps.org/prb/abstract/10.1103/PhysRevB.110.L201113}{Phys.\ Rev.\ B \textbf{110}, L201113 (2024)}.


\bibitem{ref:BBLGRTLGTTLGnew:3} Y.\ Jang, Y.\ Park, J.\ Jung, and H.\ Min, Chirality and correlations in the spontaneous spin-valley polarization of rhombohedral multilayer graphene, \href{https://journals.aps.org/prb/abstract/10.1103/PhysRevB.108.L041101}{Phys.\ Rev.\ B \textbf{108}, L041101 (2023)}.
 
\bibitem{ref:BBLGRTLGTBLGnew:1} T.\ Cea, Superconductivity induced by the intervalley Coulomb scattering in a few layers of graphene, \href{https://journals.aps.org/prb/abstract/10.1103/PhysRevB.107.L041111}{Phys.\ Rev.\ B \textbf{107}, L041111 (2023)}.

\bibitem{ref:MLGBBLGRTLGnew:1} A.\ Cr\'epieux, E.\ Pangburn, L.\ Haurie, O.\ A.\ Awoga, A.\ M.\ Black-Schaffer, N.\ Sedlmayr, C.\ P\'epin, and C.\ Bena, Superconductivity in monolayer and few-layer graphene. II. Topological edge states and Chern numbers, \href{https://journals.aps.org/prb/abstract/10.1103/PhysRevB.108.134515}{Phys.\ Rev.\ B \textbf{108}, 134515 (2023)}.


\bibitem{ref:RTLGRHLGnew:1} E.\ V.\ Bostr\"om, A.\ Fischer, J.\ B.\ Profe, J.\ Zhang, D.\ M. Kennes, and A.\ Rubio, Phonon-mediated unconventional $s$- and $f$-wave pairing superconductivity in rhombohedral stacked multilayer graphene, \href{https://www.nature.com/articles/s41524-024-01345-z}{npj Computational Materials 10, 163 (2024)}.

\bibitem{ref:RNLGnew:1} A.\ Herasymchuk, S.\ G.\ Sharapov, O.\ V.\ Yazyev, Y.\ Zhumagulov, Correlated phases in rhombohedral multilayer
graphene, \href{https://journals.aps.org/prb/abstract/10.1103/2q6v-4brs}{Phys.\ Rev.\ B {\bf 113}, 035132 (2026)}.


\bibitem{ref:Thnew:1} Y.\ Chen, M.\ S.\ Scheurer, C.\ Schrade, Intrinsic superconducting diode effect and nonreciprocal superconductivity in rhombohedral graphene multilayers, \href{https://journals.aps.org/prb/abstract/10.1103/zgnk-rw1p}{Phys.\ Rev.\ B {\bf 112}, L060505 (2025)}.

\bibitem{ref:Thnew:2} S.\ A.\ Murshed and B.\ Roy, Charge-density waves and stripes in quarter metals of graphene heterostructures, \href{https://arxiv.org/abs/2510.20816}{arXiv:2510.20816 (2025)}.



\bibitem{RMP:graphene} A.\ H.\ Castro Neto, F.\ Guinea, N.\ M.\ R.\ Peres, K.\ S.\ Novoselov, and A.\ K.\ Geim, The electronic properties of graphene, \href{https://journals.aps.org/rmp/abstract/10.1103/RevModPhys.81.109}{Rev.\ Mod.\ Phys.\ {\bf 81}, 109 (2009)}.

\bibitem{semenoff} G.\ W.\ Semenoff, Condensed-Matter simulation of a three-dimensional anomaly, \href{https://journals.aps.org/prl/abstract/10.1103/PhysRevLett.53.2449}{Phys.\ Rev.\ Lett.\ {\bf 53}, 2449 (1984)}.

\bibitem{Fierz:0} M.\ Fierz, Zur Fermischen Theorie des $\beta$-Zerfalls, \href{https://link.springer.com/article/10.1007/BF01330070}{Z.\ Physik {\bf 104}, 553 (1937)}.

\bibitem{Fierz:1} B.\ Roy, P.\ Goswami, and V.\ Juri\v{c}i\'c, Interacting Weyl fermions: Phases, phase transitions, and global phase diagram, \href{https://journals.aps.org/prb/abstract/10.1103/PhysRevB.95.201102}{Phys.\ Rev.\ B {\bf 95}, 201102 (2017)}.

\bibitem{Fierz:2} B.\ Roy and M.\ S.\ Foster, Quantum Multicriticality near the Dirac-Semimetal to Band-Insulator Critical Point in Two Dimensions: A Controlled Ascent from One Dimension, \href{https://journals.aps.org/prx/abstract/10.1103/PhysRevX.8.011049}{Phys.\ Rev.\ X {\bf 8}, 011049 (2018)}.

\bibitem{RGscheme:0} J.\ Zinn-Justin, \href{https://doi.org/10.1093/acprof:oso/9780198509233.001.0001}{\emph{Quantum Field Theory and Critical Phenomena}} (Oxford Science, Oxford, 2002).

\bibitem{RGscheme:1} A.\ L.\ Szab\'o and B.\ Roy, Extended Hubbard model in undoped and doped monolayer and bilayer graphene: Selection rules and organizing principle among competing orders, \href{https://journals.aps.org/prb/abstract/10.1103/PhysRevB.103.205135}{Phys.\ Rev.\ B {\bf 103}, 205135 (2021)}. 

\bibitem{RGscheme:2} S.\ A.\ Murshed, S.\ K.\ Das, and B.\ Roy, Superconductivity in doped planar Dirac insulators: A renormalization group study, \href{https://journals.aps.org/prb/abstract/10.1103/w6vx-375q}{Phys.\ Rev.\ B {\bf 111}, 245153 (2025)}.



\bibitem{largeNReview} M.\ Moshe and J. Zinn-Justin, Quantum field theory in the large N limit: a review, \href{https://www.sciencedirect.com/science/article/abs/pii/S0370157303002631?via%3Dihub}{Phys.\ Rept.\ {\bf 385}, 69 (2003)}.



\bibitem{kohnluttinger} W.\ Kohn and J.\ M.\ Luttinger, New Mechanism for Superconductivity, \href{https://journals.aps.org/prl/abstract/10.1103/PhysRevLett.15.524}{Phys.\ Rev.\ Lett.\ {\bf 15}, 524 (1965)}.

\bibitem{KT} J.\ M.\ Kosterlitz and D.\ J.\ Thouless, Ordering, metastability and phase transitions in two-dimensional systems, \href{https://iopscience.iop.org/article/10.1088/0022-3719/6/7/010}{J.\ Phys.\ C: Solid State Phys.\ {\bf 6}, 1181 (1973)}.


\bibitem{optgra:1} T.\ Uehlinger, G.\ Jotzu, M.\ Messer, D.\ Greif, W.\ Hofstetter, U.\ Bissbort, and T.\ Esslinger, Artificial graphene with tunable interactions, \href{https://journals.aps.org/prl/abstract/10.1103/PhysRevLett.111.185307}{Phys.\ Rev.\ Lett.\ {\bf 111}, 185307 (2013)}.

\bibitem{optgra:2} M.\ Messer, R.\ Desbuquois, T.\ Uehlinger, G.\ Jotzu, S.\ Huber, D.\ Greif, and T.\ Esslinger, Exploring Competing Density Order in the Ionic Hubbard Model with Ultracold Fermions, \href{https://journals.aps.org/prl/abstract/10.1103/PhysRevLett.115.115303}{Phys.\ Rev.\ Lett.\ {\bf 115}, 115303 (2015)}.

\bibitem{haldane} F.\ D.\ M.\ Haldane, Model for a Quantum Hall Effect without Landau Levels: Condensed-Matter Realization of the ``Parity Anomaly", \href{https://journals.aps.org/prl/abstract/10.1103/PhysRevLett.61.2015}{Phys.\ Rev.\ Lett.\ {\bf 61}, 2015 (1988)}.

\bibitem{optgra:3} G.\ Jotzu, M.\ Messer, R.\ Desbuquois, M.\ Lebrat, T.\ Uehlinger, D.\ Greif, and T.\ Esslinger, Experimental realization of the topological Haldane model with ultracold fermions, \href{https://www.nature.com/articles/nature13915}{Nature (London) {\bf 515}, 237 (2014)}.

\bibitem{dataavailability} Sk A.\ Murshed, Pair density wave in quarter metals from a repulsive fermionic interaction in graphene heterostructures: A renormalization group study, Zenodo (2026), doi:\href{https://zenodo.org/records/20028693}{10.5281/zenodo.19122266}. 

\end{thebibliography}
\end{document}